\documentclass{article}
\usepackage{spconf,amsmath,graphicx}
\usepackage{multirow}
\usepackage{subfig}
\usepackage{hyperref}
\usepackage{longtable}
\usepackage{float}
\usepackage{cite}
\usepackage{graphics}

\title{SPEAKER DIARIZATION WITH REGION PROPOSAL NETWORK}
\name{
\begin{tabular}{c}
Zili Huang$^{1}$,
Shinji Watanabe$^{1}$,
Yusuke Fujita$^{2}$,
Paola Garc\'ia$^{1}$,
Yiwen Shao$^{1}$,\\
Daniel Povey$^{1}$,
Sanjeev Khudanpur$^{1}$
\end{tabular}
}
\address{
$^{1}$ Center for Language and Speech Processing, Johns Hopkins University, USA \\
$^{2}$ Hitachi, Ltd. Research \& Development Group, Japan
}
\begin{document}
\ninept
\maketitle
\begin{abstract}

Speaker diarization is an important pre-processing step for many speech applications, and it aims to solve the ``who spoke when" problem. Although the standard diarization systems can achieve satisfactory results in various scenarios, they are composed of several independently-optimized modules and cannot deal with the overlapped speech. In this paper, we propose a novel speaker diarization method: Region Proposal Network based Speaker Diarization (RPNSD). In this method, a neural network generates overlapped speech segment proposals, and compute their speaker embeddings at the same time. Compared with standard diarization systems, RPNSD has a shorter pipeline and can handle the overlapped speech. Experimental results on three diarization datasets reveal that RPNSD achieves remarkable improvements over the state-of-the-art x-vector baseline.
\end{abstract}
\section{Introduction}
\label{sec:intro}

Speaker diarization, the process of partitioning an input audio stream into homogeneous segments according to the speaker identity\cite{reynolds2005approaches, tranter2006overview, wooters2007icsi, anguera2012speaker} (often referred as ``who spoke when"), is an important pre-processing step for many speech applications. 



\begin{figure}[!ht]
    \centering
    \includegraphics[width=1.0\linewidth]{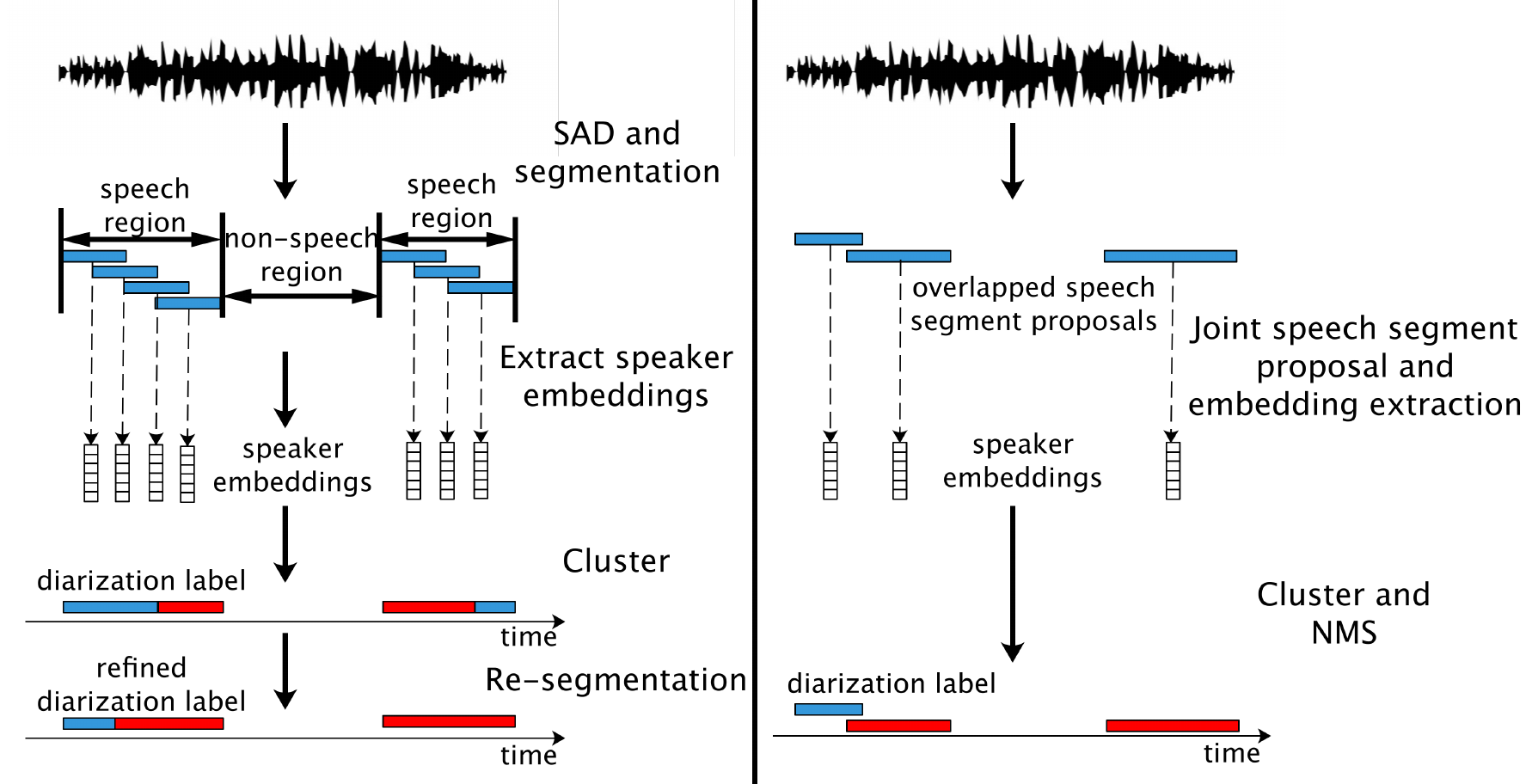}
    \caption{Pipelines of the standard diarization system (left) and the RPNSD system (right)}
    \label{fig:pipeline}
\end{figure}

As shown in Figure \ref{fig:pipeline} left, a standard diarization system\cite{sell2018diarization, diez2018but, sun2018speaker, vinals2018estimation} consists of four steps. (1) Segmentation: this step removes the non-speech portion of the audio with speech activity detection (SAD), and the speech regions are further cut into short segments. (2) Embedding extraction: in this step, a speaker embedding is extracted for each short segment. Typical speaker embeddings include i-vector\cite{kenny2007joint,dehak2010front,sell2014speaker, sell2015diarization} and deep speaker embeddings\cite{variani2014deep, heigold2016end, wan2018generalized, wang2018speaker, li2017deep, nagrani2017voxceleb, chung2018voxceleb2, snyder2017deep, snyder2018x, garcia2017speaker, sell2018diarization}. (3) Clustering: after the speaker embedding is extracted for each short segment, the segments are grouped into different clusters. Each cluster corresponds to one speaker identity. (4) Re-segmentation: this is an optional step that further refines the diarization prediction. Among the re-segmentation methods, VB re-segmentation\cite{kenny2008bayesian,diez2018speaker,sell2015diarization} is the most famous one.

Despite the successful applications in many scenarios, standard diarization systems have two major problems. (1) Many individual modules: to build a diarization system, you need a SAD model, a speaker embedding extractor, a clustering module and a re-segmentation module, all of which are optimized individually. (2) Overlap: the standard diarization system cannot handle the overlapped speech. To deal with the overlapped speech, some new modules are needed to detect and classify the overlaps, which makes the procedure even more complicated. The overlapped speech will also hurt the performance of clustering, which is the main reason standard diarization systems cannot perform well in highly overlapped scenarios \cite{fujita2019blstm}\cite{fujita2019transformer}.

Inspired by Faster R-CNN\cite{ren2015faster}, one of the best-known frameworks in object detection, we propose Region Proposal Network based Speaker Diarization (RPNSD). As shown in Figure \ref{fig:pipeline} right, in this method, we combine the segmentation, embedding extraction and re-segmentation into one stage. The segment boundaries and speaker embeddings are jointly optimized in one neural network. After the speech segments and corresponding speaker embeddings are extracted, we only need to cluster the segments and apply non-maximum suppression (NMS) to get the diarization prediction, which is much more convenient than the standard diarization system. In addition to that, since the speech segment proposals overlap with each other, our framework solves the overlap problem in a natural and elegant way.

The experimental results on Switchboard, CALLHOME and simulated mixtures reveal that our framework achieves significant and consistent improvements over the state-of-the-art x-vector baseline, and a great portion of the improvements come from successfully detecting the overlapped speech regions. Our code is available at \url{https://github.com/HuangZiliAndy/RPNSD}.

\section{Methodology}
\label{sec:method}

In this section, we will introduce our framework in details. Our framework aims to solve the speaker diarization problem and it consists of two steps. (1) Joint speech segment proposal and speaker embedding extraction. (2) Post-processing. In the first step, we predict the boundary of speech segments and extract speaker embeddings with one neural network. In the second step, we perform clustering and apply NMS to get diarization predictions.


\subsection{Joint Speech Segment Proposal and Embedding Extraction}

\begin{figure}[!ht]
    \centering
    \includegraphics[width=1.0\linewidth]{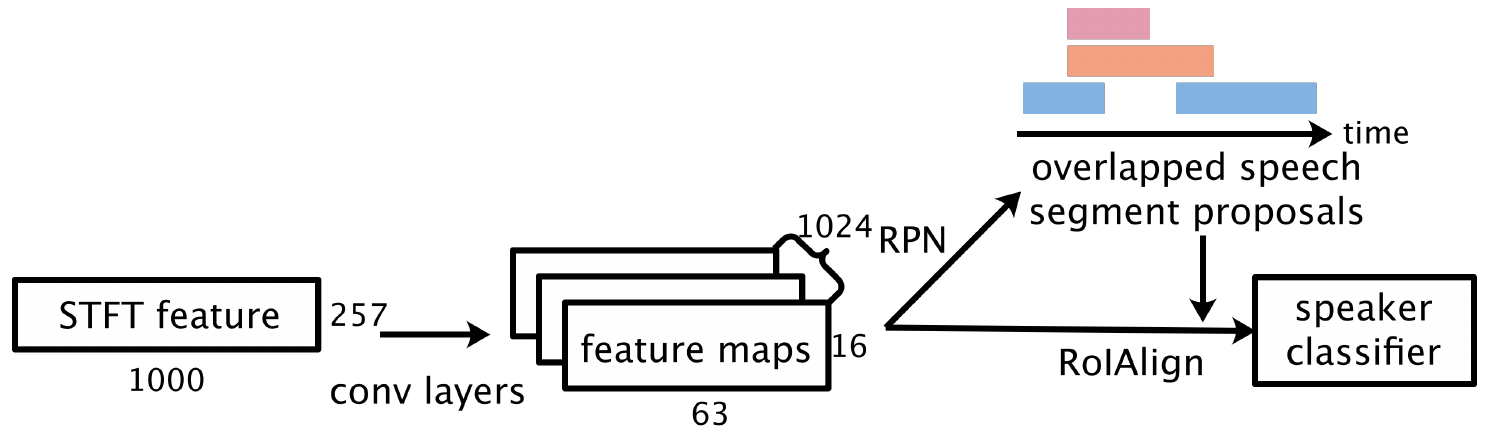}
    \caption{Procedure of the first step: joint speech segment proposal and speaker embedding extraction}
    \label{fig:step1}
\end{figure}

The overall procedure of the first step is shown in Figure \ref{fig:step1}. Given an audio input, we first extract acoustic features\footnote{We  experiment on 8kHz telephone data and we choose the STFT feature with frame size 512 and frame shift 80. During training we segment the audios into 10s chunks, so the feature shape of each chunk is (257, 1000).} and feed them into convolution layers to obtain the feature maps. Then a Region Proposal Network (RPN) will generate many overlapped speech segment proposals \cite{kao2018r} and predict their confidence scores. After that, the deep features corresponding to the speech segment proposals are pooled into fixed-size representations. Finally, we perform speaker classification and boundary refinement on the top of the representations.

\subsubsection{Region Proposal Network (RPN)}
\label{subsubsection:RPN}
The RPN\cite{ren2015faster} is the key component of our framework. It takes the feature maps as the input and outputs the region proposals. The original RPN generates 2-d region proposals while our RPN generates 1-d speech segment proposals.  In our framework, the RPN takes the feature maps as the input\footnote{The size of the feature maps is (1024, 16, 63). There are 63 timesteps and each timestep corresponds 16 frames of speech.} and predicts speech segment proposals. Similar to brute-force search, the RPN will consider every timestep as a possible center and expand several anchors with pre-defined sizes from it. In our system, we use 9 anchors with the size of $\{1, 2, 4, 8, 16, 24, 32, 48, 64\}$, which covers the speech segments from 16 to 1024 frames. Meanwhile, the RPN will also predict scores and refine boundaries for each speech segment proposal with convolution layers. Among the $63 \times 9 = 567$ (63 timesteps and 9 anchors per timestep) speech segment proposals, we first filter out the speech segment proposals with low confidence scores and then further remove highly overlapped segments with NMS. In the end, we keep 100 high-quality speech segment proposals after NMS during training and 50 during evaluation.

\subsubsection{RoIAlign}
After the RPN predicts the speech segment proposals, we extract corresponding regions from the feature maps as the deep features for each segment. Since the sizes of speech segment proposals vary a lot, we need RoIAlign\cite{he2017mask} to pool the features into fixed dimension. Suppose we want to pool the $D \times T$ speech segment proposal ($D$ is the feature dimension and $T$ is the unfixed timestep) into a fixed representation, the proposed region is first divided into $N \times N$ ($N = 7$) RoI bins. Then we uniformly sample four locations in each RoI bin and use bilinear interpolation to compute the values of them. The result is aggregated using average pooling. With the pooled feature of fixed dimension, we can perform speaker classification and boundary refinement for each speech segment proposal.

\subsubsection{Loss Function}
The training loss consists of five parts and is formulated as
\begin{equation}
\label{eq:total}
L = L_\mathrm{rpn\_cls} + L_\mathrm{rpn\_reg} + L_\mathrm{rcnn\_cls} + L_\mathrm{rcnn\_reg} + \alpha \cdot L_\mathrm{spk\_cls}
\end{equation} 
In equation \ref{eq:total}, $L_\mathrm{rpn\_cls}$ and $L_\mathrm{rcnn\_cls}$ are binary cross-entropy loss to classify foreground/background (fg/bg), which is formulated as
\begin{equation}
L_\mathrm{cls}(p_i, p_i^{\ast})=-(p_i^{\ast}\log(p_i) + (1-p_i^{\ast})\log(1-p_i))
\end{equation}
where $p_i$ is the probability that the speech segment $i$ is foreground and $p_i^{\ast}$ is the ground truth label. Whether a segment is fg or bg is determined by the Intersection-over-Union (IoU) overlap with the ground-truth segments.
$L_\mathrm{rpn\_reg}$ and $L_\mathrm{rcnn\_reg}$ are regression loss to refine the speech segment boundaries, which are formulated as
\begin{equation}
L_\mathrm{reg}(\mathbf{t}_i, \mathbf{t}_i^{\ast}) = R(\mathbf{t}_i - \mathbf{t}_i^{\ast})
\end{equation}
where $\mathbf{t}_i$ and $\mathbf{t}_i^{\ast}$ are the coordinates of predicted segments and ground truth segments respectively, and $R$ is the smooth L1 loss function in \cite{girshick2015fast}. The coordinates $\mathbf{t}_i$ and $\mathbf{t}_i^{\ast}$ are defined as follows.

\begin{align}
    \mathbf{t}_i &= [(x - x_a) / w_a, \log(w / w_a)] \\
    \mathbf{t}_i^{\ast} &= [(x^{\ast} - x_a) / w_a, \log(w^{\ast} / w_a)]
\end{align}
where $x$ and $w$ denote the center position and length of the segment. $x$, $x_a$ and $x^{\ast}$ represent the center positions for the predicted segment, anchor and ground truth segment respectively (likewise for $w$). $L_\mathrm{spk\_cls}$ is the cross-entropy loss to classify the segment’s speaker identity, which is defined as
\begin{equation}
L_\mathrm{spk\_cls}(\mathbf{s}_i, \mathbf{s}_i^{\ast}) = -\mathbf{s}_i^{\ast} \cdot \log{(\mathbf{s}_i)} 
\end{equation}
where $\mathbf{s}_i$ is the predicted probability distribution over all speakers in the training set and $\mathbf{s}_i^{\ast}$ is the ground truth one-hot speaker label. $L_\mathrm{spk\_cls}$ is scaled with a weight factor $\alpha$.

Among the loss components, $L_\mathrm{rpn\_cls}$ and $L_\mathrm{rpn\_reg}$ are used to train the RPN. We adopt the same strategy as \cite{ren2015faster}, and sample 128 from 567 initial speech segment proposals to compute $L_\mathrm{rpn\_cls}$ and $L_\mathrm{rpn\_reg}$. The segment proposals having an IoU overlap higher than 0.7 with any ground-truth segments are labeled as fg while the segment proposals with an IoU overlap lower than 0.3 for all ground-truth segments are labeled as bg. $L_\mathrm{rpn\_reg}$ is calculated only for the fg. $L_\mathrm{rcnn\_cls}$ and $L_\mathrm{rcnn\_reg}$ have the exactly same form but are calculated with different samples. We sample 64 from the 100 high-quality speech segments mentioned in section \ref{subsubsection:RPN} to compute $L_\mathrm{rcnn\_cls}$ and $L_\mathrm{rcnn\_reg}$. $L_\mathrm{spk\_cls}$ is also calculated with the 64 samples, and it ensures that we extract discriminative embeddings from the model. 


\subsection{Post-processing}
In RPNSD, the input of the first step is an audio and the output includes: (1) the speech segment proposals, (2) the probability of fg/bg and (3) the speaker embedding for each segment proposal. In the second step, we perform post-processing to get the diarization prediction. The whole process contains three steps. 
\begin{enumerate}
    \item Remove the speech segment proposals whose fg probability is lower than a threshold $\gamma$. ($\gamma = 0.5$ in our experiment)
    \item Clustering: Group the remaining speech segment proposals into clusters. (We use K-means in our experiment)
    \item Apply NMS (NMS threshold = 0.3) for segments in the same cluster to remove the highly overlapped segment proposals.
\end{enumerate}

\section{Experiments}
\label{sec:experiments}
\subsection{Datasets and Evaluation Metrics}
\label{subsec:datasets}
\subsubsection{Datasets}
We train our systems on two datasets (Mixer 6 + SRE + SWBD and Simulated TRAIN) and evaluate on three datasets (Switchboard, CALLHOME and Simulated DEV) to verify the effectiveness of our framework. The dataset statistics are shown in Table \ref{tab:data statistics}. The overlap ratio is defined as $overlap{\;}ratio = \frac{t_{ spk \geq 2}}{t_{spk \geq 1}}$, where $t_{ spk \geq n}$ denotes the total time of speech regions with more than $n$ speakers. Since end-to-end systems are usually data hungry and require massive training data to generalize better, we come up with two methods to create huge amount of diarization data. (1) Use public telephone conversation datasets (Mixer 6 + SRE + SWBD). (2) Use speech data of different speakers to create synthetic diarization datasets (Simulated TRAIN). Detailed introductions for each dataset are as follows.

\begin{table}[!ht]
    \centering
    \begin{tabular}{c c c c} \hline
        &  \# utts & avg. dur & overlap ratio \\
        & & (sec) & (\%)\\\hline
        \textbf{Train sets} & & &  \\
        Mixer 6 + SRE + SWBD & 29,697 & 348.1 &  5.0 \\
        Simulated TRAIN($\beta = 2$) & 100,000 & 87.6 & 34.4 \\\hline
        \textbf{Test sets} & & &  \\
        CALLHOME & 499 & 124.5 & 16.9 \\
        SWBD DEV & 99 & 304.6 & 5.2 \\
        SWBD TEST & 100 & 312.0 & 5.8 \\
        Simulated DEV($\beta = 2$) & 500 & 87.3 & 34.4 \\
        Simulated DEV($\beta = 3$) & 500 & 103.8 & 27.2 \\
        Simulated DEV($\beta = 5$) & 500 & 137.1 & 19.5 \\\hline
    \end{tabular}
    \caption{Dataset statistics}
    \label{tab:data statistics}
\end{table}

The Mixer 6 + SRE + SWBD dataset includes Mixer 6, SRE04-10, Switchboard-2 Phase I-III and Switchboard Cellular Part 1, 2, and the majority of the dataset are 8kHz telephone conversations. For speaker recognition, we usually use single channel audios that contain only one person. While in our experiment, we sum up both channels to create a large diarization dataset. The ground truth diarization label is generated by applying SAD on single channels.\footnote{In all experiments of this paper, we use the TDNN SAD model (\url{http://kaldi-asr.org/models/m4}) trained on the Fisher corpus.} We also used the same data augmentation technique as \cite{snyder2018x} and the train sets are augmented with music, noise and reverberation from the MUSAN\cite{snyder2015musan} and the RIR\cite{ko2017study} dataset. The augmented train set contains 10,574 hours of speech.

SWBD DEV and SWBD TEST are sampled from the SWBD dataset (We exclude these audios from Mixer6 + SRE + SWBD). They contain around 100 5-minute audios and share no common speaker with the train set. Like Mixer 6 + SRE + SWBD, the overlap ratio of SWBD DEV and SWBD TEST is quite low. We create these two datasets to evaluate the system performance on similar data.

The CALLHOME dataset is one of the best-known benchmarks for speaker diarization. As one part of 2000 NIST Speaker Recognition Evaluation (LDC2001S97), the CALLHOME dataset contains 500 audios in 6 languages including Arabic, English, German, Japanese, Mandarin, and Spanish. The number of speakers in each audio ranges from 2 to 7.

We also use synthetic datasets (same as \cite{fujita2019blstm, fujita2019transformer}) to evaluate RPNSD's performance on highly overlapped speech. The simulated mixtures are made by placing two speakers' speech segments in a single audio file. The human voices are taken from SRE and SWBD, and we use the same data augmentation technique as \cite{snyder2018x}. The parameter $\beta$ is the average length of silence intervals between segments of a single speaker, and a larger $\beta$ results in less overlap. In our experiment, we generate a large dataset with $\beta = 2$ for training and three datasets with $\beta = 2, 3, 5$ for evaluation. The training set and test set share no common speaker.

\subsubsection{Evaluation Metrics}
We evaluate different systems with Diarization Error Rate (DER). The DER includes Miss Error (speech predicted as non-speech or two speaker mixture predicted as one speaker etc.), False Alarm Error (non-speech predicted as speech or single speaker speech predicted as multiple speaker etc.) and Confusion Error (one speaker predicted as another). Many previous studies\cite{snyder2017deep, zhang2019fully} ignore the overlapped regions and use 0.25s collar for evaluation. While in our study, we score the overlapped regions and report the DER with different collars.

\subsection{Baseline}
We follow Kaldi's CALLHOME diarization V2 recipe\cite{povey2011kaldi} to build baselines. The recipe uses oracle SAD labels which are not available in real situations, so we first use a TDNN SAD model to detect the speech segments. Then the speech segments are cut into 1.5s chunks with 0.75s overlap, and x-vectors are extracted for each segment. After that, we apply Agglomerative Hierarchical Clustering (AHC) to group segments into different clusters, and the similarity matrix is based on PLDA\cite{prince2007probabilistic} scoring. We also apply VB re-segmentation for CALLHOME experiments.

\subsection{Experimental Settings}
We use ResNet-101 as the network architecture and Stochastic Gradient Descent (SGD) as the optimizer \footnote{We refer the PyTorch implementation of Faster R-CNN in \cite{jjfaster2rcnn}.}. We start training with a learning rate of $0.01$ and it decays twice to $0.0001$. The batch size is set as 8 and we train our model on NVidia GTX 1080 Ti for around 4 days. The scaling factor $\alpha$ in equation \ref{eq:total} is set to $1.0$ for training. During adaptation, we use a learning rate of $4 \cdot 10^{-5}$ and $\alpha$ is set to $0.1$. The speaker embedding dimension is 128.

\subsection{Experimental Results}
\subsubsection{Experiments on Switchboard}
\label{subsec:exp_swbd}

\begin{table}[!ht]
    \centering
    \begin{tabular}{c c c c c} \hline
        \multirow{2}{4em}{Dataset} & \multirow{2}{4em}{System} & DER(\%) & DER(\%) & DER(\%) \\
        & & c=0s & c=0.1s & c=0.25s \\\hline\hline
        \multirow{2}{4em}{SWBD DEV} & x-vector & 15.39 & 9.51 & 4.66 \\
        & RPNSD & \textbf{9.18} & \textbf{4.09} & \textbf{2.50} \\\hline
        
        \multirow{2}{4em}{SWBD TEST} & x-vector & 15.08 & 9.36 & 4.42 \\
        & RPNSD & \textbf{9.09} & \textbf{4.14} & \textbf{2.55} \\\hline
    \end{tabular}
    \caption{DERs (\%) on SWBD DEV and SWBD TEST with different collars, the overlapped speech is also scored.}
    \label{tab:swbd}
\end{table}

\begin{table*}[t]
    \centering
    \begin{tabular}{c | c | c | c c c | c c c} \hline
        \multirow{2}{4em}{System} & \multirow{2}{4em}{SAD} & \multirow{2}{4em}{Cluster} & \multicolumn{3}{c|}{DER(\%) Score Overlap} & \multicolumn{3}{c}{DER(\%) Not Score Overlap} \\
        & & & c=0s & c=0.1s & c=0.25s & c=0s & c=0.1s & c=0.25s\\\hline\hline
        x-vector & oracle & AHC with threshold & 25.07 & 21.75 & 17.57 & 12.88 & 10.60 & 8.02 \\
        x-vector & oracle & AHC with oracle \# spk & 24.13 & 20.76 & 16.54 & 11.63 & 9.33 & 6.73 \\
        x-vector (+VB) & oracle & AHC with threshold & 23.47 & 19.89 & 16.38 & 10.68 & 8.15	& 6.51 \\
        x-vector (+VB) & oracle & AHC with oracle \# spk & 22.12 & 18.47 & 14.91 & 9.11& 6.53 & 4.90 \\\hline
        x-vector & TDNN SAD & AHC with threshold & 32.63 & 26.62 & 20.71 & 23.23 & 16.85 & 11.70 \\
        x-vector & TDNN SAD & AHC with oracle \# spk & 32.20 & 26.13 & 20.14 & 22.53 & 16.10 & 10.90 \\
        x-vector (+VB) & TDNN SAD & AHC with threshold & 30.44 & 24.69 & 19.51 & 20.06 & 14.17 & 10.09 \\
        x-vector (+VB) & TDNN SAD & AHC with oracle \# spk & 29.54 & 23.77 & 18.61 & \textbf{19.06} & \textbf{13.18} & \textbf{9.14} \\
        RPNSD & / & K-means with oracle \# spk & \textbf{25.46} & \textbf{20.41} & \textbf{17.06} & 21.39 & 15.35 & 11.81 \\\hline
    \end{tabular}
    \caption{DERs (\%) on CALLHOME with different scoring options}
    \label{tab:CALLHOME}
\end{table*}

In this experiment, we train RPNSD on Mixer 6 + SRE + SWBD and use Kaldi's x-vector model for CALLHOME as the baseline.\footnote{We also train a x-vector model on single channel data of Mixer 6 + SRE + SWBD as a fair comparison but the performance is slightly worse than Kaldi's diarization model (\url{http://kaldi-asr.org/models/m6}).} As shown in Table \ref{tab:swbd}, RPNSD significantly reduces the DER from $15.39\%$ to $9.18\%$ on SWBD DEV and $15.08\%$ to $9.09\%$ on SWBD TEST. On SWBD TEST, the DER composition of the x-vector baseline is $8.9\%$ (Miss) + $1.1\%$ (False Alarm) + $5.0\%$ (Speaker Confusion) = $15.08\%$ (with $2.8\%$ Miss and $0.9\%$ False Alarm for SAD). For RPNSD, the DER composition is $4.0\%$ (Miss) + $4.8\%$ (False Alarm) + $0.3\%$ (Speaker Confusion) = $9.09\%$ (with $2.0\%$ Miss and $2.2\%$ False Alarm for SAD).

Since RPNSD can handle the overlapped speech, the miss error decreases from $8.9\%$ to $4.0\%$. As a cost, the false alarm error increases from $1.1\%$ to $4.8\%$. Surprisingly, the speaker confusion decreases largely from $5.0\%$ to $0.3\%$. There might be two reasons for this. (1) Instead of making decisions on short segments, RPNSD makes use of longer context and extracts more discriminative speaker embeddings. (2) The training and testing condition are more matched for RPNSD. Instead of training on single speaker data, we are training on ``diarization data" and testing on ``diarization data".  

\subsubsection{Experiments on CALLHOME}

The CALLHOME corpus is one of the best-known benchmarks for speaker diarization. Since the CALLHOME corpus is quite small (with 17 hours of speech) and doesn't specify dev/test splits, we follow the ``pre-train and adapt" procedure and perform a 5-fold cross validation on this dataset. We use the model in section \ref{subsec:exp_swbd} as the pre-trained model, adapt it on $4/5$ of CALLHOME data and evaluate on the rest $1/5$. Since our model does not use any segment boundary information, it is unfair to compare it with x-vector systems using the oracle SAD label. Therefore we compare it with x-vector systems using TDNN SAD. As shown in Table $\ref{tab:CALLHOME}$, our system achieves better results than x-vector systems with and w/o VB re-segmentation. It largely reduces the DER from $32.30\%$ (or $29.54\%$ after VB re-segmentation) to $25.46\%$. The detailed DER breakdown is shown in Table \ref{tab:CALLHOME_DER}. Due to the ability to handle overlapped speech, RPNSD largely reduces the Miss Error from $18.6\%$ to $12.8\%$. As a cost, the False Alarm Error increases from $5.1\%$ to $7.5\%$. The Confusion Error of RPNSD is also lower than x-vector and x-vector (+VB).

The DER result of RPNSD ($25.46\%$) is even close to the x-vector system using the oracle SAD label ($24.13\%$). If the oracle SAD label is used, the DER of RPNSD system must be lower than $25.46 - 3.2 = 22.26\%$\footnote{This is because we can easily remove the False Alarm SAD error by labeling them as silence. It is more difficult to handle the Miss SAD error in this framework, but we can further reduce the DER for sure.}, which is better than the x-vector system ($24.13\%$) and quite close to x-vector (+VB) ($22.12\%$). 

\begin{table}[!ht]
    \centering
    \begin{tabular}{c | c | c c c | c c} \hline
        & & \multicolumn{3}{c|}{DER breakdown} & \multicolumn{2}{c}{SAD error}\\
        System & DER & MI & FA & CF & MI & FA \\\hline\hline
        x-vector & 32.20 & 18.6 & 5.1 & 8.6 & 4.2 & 5.3 \\
        x-vector (+VB) & 29.54 & 18.6 & 5.1 & 5.9 & 4.2 & 5.3 \\
        RPNSD & 25.46 & 12.8 & 7.5 & 5.2 & 5.2 & 3.2 \\\hline
    \end{tabular}
    \caption{The DER composition of different diarization systems on CALLHOME dataset. The DER includes Miss Error (MI), False Alarm Error (FA), and Confusion Error (CF). The SAD error includes Miss (MI) and False Alarm (FA).}
    \label{tab:CALLHOME_DER}
\end{table}

\subsubsection{Experiments on Simulated Mixtures}
According to our experience, standard diarization systems fail to perform well on highly overlapped speech. Therefore we design experiments on simulated mixtures to evaluate the system performance on overlapped scenarios. As shown in Table \ref{tab:simulated}, RPNSD achieves much lower DER than i-vector and x-vector systems. Compared with permutation-free loss based end-to-end systems\cite{fujita2019blstm, fujita2019transformer}, the performance of RPNSD is better than BLSTM-EEND but worse than SA-EEND. However, unlike these two systems, RPNSD does not have any constraint on the number of speakers. 


\begin{table}[!ht]
    \centering
    \begin{tabular}{c c c c} \hline
        \multirow{2}{4em}{System} & \multicolumn{3}{c}{Simulated} \\ 
        & $\beta=2$ & $\beta=3$ & $\beta=5$ \\\hline\hline
        i-vector & 33.74 & 30.93 & 25.96 \\
        x-vector & 28.77 & 24.46 & 19.78 \\
        BLSTM-EEND & 12.28 & 14.36 & 19.69 \\
        SA-EEND & \textbf{7.91} & \textbf{8.51} & \textbf{9.51} \\
        RPNSD & 9.30 & 11.57 & 14.55 \\\hline
    \end{tabular}
    \caption{DERs (\%) on simulated mixtures with 0.25s collar, the overlapped speech is also scored.}
    \label{tab:simulated}
\end{table}

\section{Conclusion}
\label{sec:conclusion}
In this paper, we propose a novel speaker diarization system RPNSD. Taken an audio as the input, the model predicts speech segment proposals and speaker embeddings at the same time. With some simple post-processing (clustering and NMS), we can get the diarization prediction, which is much more convenient than the standard process. In addition to that, the RPNSD system solves the overlapping problem in an elegant way. Our experimental results on Switchboard, CALLHOME and synthetic mixtures reveal that the improvements of the RPNSD system are obvious and consistent.


\bibliographystyle{IEEEbib}
\bibliography{strings,refs}

\end{document}